\documentclass{llncs}%
\usepackage[pdftex]{graphicx}%
\usepackage{amsmath}%
\usepackage{amssymb}%
\usepackage{url}%
\usepackage[all]{xy}%
\usepackage{caption}%
\usepackage{subcaption}%
\usepackage{paralist}%
\usepackage{color}%
\usepackage{enumitem}%
\usepackage[compact]{titlesec}%
\usepackage{algorithm}%
\usepackage{algpseudocode}%
\usepackage{tabto}%
\newcommand{\pjs}[1]{\marginpar{\sc pjs}\textcolor{blue}{#1}}%
\newcommand{\nd}[1]{\marginpar{\sc nd}\textcolor{blue}{#1}}%
\newcommand{\ignore}[1]{}%
\newcommand\abs[1]{\vert#1\vert}%
\newcommand\linear{\textit{linear}}%
\newcommand\chuffed{\textit{Chuffed}}%
\newcommand\cplex{\textit{CPLEX}}%
\newcommand\scip{\textit{SCIP}}%
\newcommand\toolbar{\textit{toolbar}}%
\newcommand\toulbar{\textit{toulbar2}}%
\newcommand\akmaxsat{\textit{akmaxsat}}%
\newcommand\maxsatz{\textit{MaxSatz}}%
\newcommand\minizinc{\textit{Minizinc}}%
\newcommand\false{\textit{false}}%
\newcommand\true{\textit{true}}%
\newcommand\lit[1]{[#1]}
\newcommand\model[1]{\textsc{#1}}%
\newcommand\T{\rule{0pt}{2.2ex}}%
\setlist{topsep=1pt plus 2pt}%
\setlist[trivlist]{topsep=6pt plus 2pt minus 1pt}%
\titlespacing*{\section} {0pt}{3ex plus .5ex minus .2ex}{2ex plus .2ex}%
\titlespacing*{\subsection} {0pt}{2.25ex plus .5ex minus .2ex}{1.5ex plus .2ex}%
\title{Unsatisfiable Cores and Lower Bounding for Constraint Programming}%
\author{Nicholas Downing  \and Thibaut Feydy \and Peter J. Stuckey}%
\institute{%
National ICT Australia\thanks{NICTA is funded by the Australian Government as represented by the Department of Broadband, Communications and the Digital Economy and the Australian Research Council.}\ %
and the University of Melbourne, Victoria, Australia \\
\email{\{ndowning,tfeydy,pjs\}@csse.unimelb.edu.au}%
}%
\algrenewcommand{\algorithmiccomment}[1]{\% #1}%
\begin{document}%
%
\abovedisplayskip 6pt plus 2pt minus 1pt%
\abovedisplayshortskip 0pt plus 1pt%
\belowdisplayskip 6pt plus 2pt minus 1pt%
\belowdisplayshortskip 0pt plus 1pt%
\maketitle%
\begin{abstract}%
Constraint Programming (CP) solvers typically tackle optimization problems
by repeatedly finding solutions to a problem while placing tighter and tighter
bounds on the solution cost.  This approach is somewhat naive, especially
for soft-constraint optimization 
problems in which the soft constraints are mostly satisfied.  
Unsatisfiable-core approaches to solving soft constraint problems in SAT 
(e.g. MAXSAT)  
force all soft constraints to
be hard initially. When solving fails they return an unsatisfiable core, as a set of soft constraints
that cannot hold simultaneously. 
These are reverted to soft and solving continues.
Since lazy clause generation solvers can also return unsatisfiable cores
we can adapt this approach to constraint programming.
We adapt the original MAXSAT unsatisfiable core solving approach 
to be usable for constraint programming and
define a number of extensions.  
Experimental results show that our methods are beneficial on a
broad class of CP-optimization benchmarks involving soft constraints,
cardinality or preferences.
\end{abstract}%
\section{Introduction}\label{sec:introduction}%
Earlier work on unsatisfiable cores for Maximum Satisfiability (MAXSAT) has shown that it is advantageous to consider soft constraints to be hard constraints initially, solve the problem using a modern SAT solver, and use the resulting evidence of infeasibility to see which (temporarily hard) constraints are conflicting with each other, and soften them again only as necessary~\cite{fu}.

In this paper we extend the unsatisfiable cores algorithm from MAXSAT to Constraint Programming (CP).  CP handles soft-constraint problems as minimization problems where the objective is a count of violations, the counts being derived from either reified primitive constraints (whose enforcement is controlled by an auxiliary variable) or soft global constraints (for examples of soft global constraints and their propagation algorithms see Van Hoeve~\cite{vanhoeve}).

One reason to expect that unsatisfiable cores will help in CP is that propagation solving relies on eliminating impossible (partial) solutions, but unfortunately when most constraints are soft then most solutions cannot be ruled out definitively and so propagation has little effect.  Making as many constraints as possible hard, should improve the propagation behaviour.  Conversely, for the approach to be successful the solver needs to be able to prove infeasibility, if this is easy to do repeatedly then unsatisfiable cores will be highly effective, but if it requires a lot of search then it should be put off for as long as possible!

We work in the context of a Lazy Clause Generation (LCG) solver, because the
LCG solver can easily `explain' why failures occurred, which is useful
because it tells us which (temporarily) hard constraints should be made soft
again.  LCG is a hybrid approach to CP that uses a 
traditional `propagation and search' constraint solver as the outer layer which guides the solution process, plus an inner layer which lazily decomposes CP to Boolean satisfiability (SAT) and applies learning SAT solver technology to reduce search~\cite{moskewicz,ohrimenko}.

The contributions of this paper are:
\begin{itemize}%
\item We translate the basic unsatisfiable core approach of
  SAT to CP solving.
\item We extend the basic unsatisfiable core approach to a nested version which
  more aggressively makes soft constraints hard.
\item We discuss how we can use the unsatisfiable cores generated to improve
  the estimation of the objective function in CP search.
\item We give experiments showing that for some CP optimization 
problems the
  unsatisfiable-core approach is significantly better than branch and bound.
\end{itemize}%
\section{Lazy Clause Generation}\label{sec:lazyclausegeneration}%
We give a brief description of propagation-based solving and LCG,
for more details see~\cite{ohrimenko}.
We consider problems consisting
of constraints $\mathbf{C}$ over integer variables $x_1$, $\ldots$, $x_n$,
each with a given finite domain $D_\text{orig}(x_i)$.  
A feasible solution is a valuation $\theta$ to the variables, which
satisfies all constraints $\mathbf{C}$, and lies in the domain
$\mathbf{D}_\text{orig} = D_\text{orig}(x_1) \times \ldots \times
D_\text{orig}(x_n)$, i.e.~$\theta(x_i) \in D_\text{orig}(x_i)$.

A propagation solver maintains a domain restriction $D(x_i) \subseteq
D_\text{orig}(x_i)$ for each variable and considers only solutions that lie
within $\mathbf{D} = D(x_1) \times \ldots \times D(x_n)$.  Solving interleaves
propagation, which repeatedly applies propagators to remove unsupported
values, and search which splits the domain of some variable and considers
the resulting sub-problems.  This continues until all variables are fixed
(success) or failure is detected (backtrack and try another subproblem).
A \emph{singleton domain} $D$ where all variables are fixed 
corresponds to a valuation
$\theta_D$ where $\theta_D(x_i) = v_i$ when $D(x_i) = \{v_i\}$, $i \in 1..n$.

Lazy clause generation is implemented by introducing Boolean variables for
each potential value of a CP variable, named $\lit{x_i = j}$, and
for each bound, $\lit{x_i \ge j}$.  Negating them gives $\lit{x_i \ne j}$ and
$\lit{x_i \le j - 1}$.  Fixing such a \emph{literal} modifies $D(x_i)$ to
make the corresponding fact true, and vice versa. Hence the literals give an
alternate Boolean representation of the domain, which supports reasoning.
Lazy clause generation makes use of \emph{clauses} to record nogoods.
A clause is a disjunction of literals, which we will often
treat as a set of literals.\ignore{We shall also write clauses as implications
in the form $L \rightarrow l$, where $L$ is a set of literals,
as notation for the clause $l \vee \bigvee_{m \in L} \neg m$.}

\begin{algorithm}%
\begin{algorithmic}[1]%
\Function{LCG}{$\mathbf{C}, \mathbf{D}, \mathbf{c}, \mathbf{y}$}\tabto{17em}\Comment{initial constraints, domains and objective}
  \State $S \leftarrow []$\tabto{17em}\Comment{empty stack of domains per decision level}
  \State $\theta \leftarrow \textit{none}$\tabto{17em}\Comment{best solution found so far, initially none}
  \Loop
    \State \Comment{the call below updates the implication graph and $N$, not shown explicitly}
    \State $\mathbf{D} \leftarrow \textproc{Propagate}(\mathbf{C}, \mathbf{D})$
    \If{$\mathbf{D} = \emptyset$}\tabto{17em}
      \State \Comment{failure, nogood is $N \rightarrow \false$ where $\textit{decision\_level}(l) > 0$ for all $l \in N$}
      \If{$N = \emptyset$}\tabto{17em}\Comment{conflict occurred at level 0}
        \State \Return{$\theta$}\tabto{17em}\Comment{no further improvement possible}
      \Else
        \State $(L, M) \leftarrow \textproc{Analyze}(N)$\tabto{17em}\Comment{make 1UIP nogood $L \rightarrow l$ where $M = \{l\}$}
        \State pop $S$ until reaching highest decision level of literals in $L$ or 0\label{lin:backtracklevel}
        \State pop $\mathbf{D}$ from $S$\tabto{17em}\Comment{backjump}
        \State $\mathbf{C} \leftarrow \textproc{Learn}(\mathbf{C}, L, M)$\tabto{17em}\Comment{add redundant constraint to problem}\label{lin:learn}
      \EndIf
    \ElsIf{$\mathbf{D}$ is a singleton domain}
      \State \Comment{found solution, record it and restart with tighter objective constraint}
      \State $\theta \leftarrow \theta_D$
      \State pop $S$ until reaching decision level 0
      \State pop $\mathbf{D}$ from $S$\tabto{17em}
      \State $\mathbf{C} \leftarrow \mathbf{C} \cup \{\mathbf{c}^T
      \mathbf{y} < \mathbf{c}^T \theta(\mathbf{y}) \}$
    \Else
      \State \Comment{reached a fixed point, execute the user's programmed search strategy}
      \State push $\mathbf{D}$ onto $S$
      \State $\mathbf{D} \leftarrow \textproc{Decide}(\mathbf{D})$
    \EndIf
  \EndLoop
\EndFunction
\vspace*{2em plus 3em minus 1em}%
\Function{Analyze}{$N$}
  \State $\textit{conflict\_level} \leftarrow \max_{n \in N} \textit{decision\_level}(n)$
  \While{there are multiple $n \in N$ with $\textit{decision\_level}(n) = \textit{conflict\_level}$}
    \State let $L \rightarrow l$ be the most recent unprocessed propagation
    at $\textit{conflict\_level}$\label{lin:unprocessed}
   \State \algorithmicif{} no such propagations remain unprocessed \algorithmicthen{} break \algorithmicend{} \algorithmicif{} \label{lin:exit}
    \State \algorithmicif{} $l \in N$ \algorithmicthen{} $N \leftarrow (N - \{l\}) \cup L$ \algorithmicend{} \algorithmicif{}
  \EndWhile
  \State \Return{$(\{n: n \in N, \textit{decision\_level}(n) < \textit{conflict\_level}\},$}
  \Statex \hspace*{6em}$\{\neg n: n \in N, \textit{decision\_level}(n) = \textit{conflict\_level}\})$
\EndFunction
\vspace*{2em plus 3em minus 1em}%
\Function{Learn}{$\mathbf{C}, L, \{l\}$}
  \State \Return{$\mathbf{C} \cup \{L \rightarrow l\}$}
\EndFunction
\end{algorithmic}%
\caption{CP branch-and-bound with clause learning and backjumping}\label{alg:dpll}%
\end{algorithm}%
The high-level solving algorithm \textproc{LCG}, including propagation,
search, and nogood generation, is shown as Algorithm~\ref{alg:dpll}.  It is
a standard 
CP branch-and-bound search, except that propagation (\textsc{Propagate}) 
must return a nogood $N$ as shown, explaining any failures that are detected by propagation.  Propagation must also record an \emph{implication graph} showing the reasons for each propagation step, and for each literal which is fixed, its \emph{decision level} as the value of $\abs{S}$ at the time of fixing.  \emph{Conflict analysis} derives new redundant constraints to avoid repeated search, and, as a side effect, modifies the backtracking procedure to \emph{backjump} or restart solving at an appropriate point close to the failure~\cite{moskewicz}.

The \textproc{Analyze} procedure reduces the information from the failure
nogood $N \rightarrow \false$, and the implication graph, into a 1UIP nogood,
which can be learnt as a new redundant constraint.
It considers propagations at the \emph{conflict level}, which is the highest
level of any literal in $N$.  Working in reverse chronological order, for
each propagation $L \rightarrow l$, where the propagated literal $l$ occurs
in $N$, this literal is replaced by its reason giving $(N - \{l\}) \cup L$.
The process stops when there is at most one literal in $N$ whose decision
level is the conflict level, leaving a clause which
propagates to fix that literal to its opposite value.

%
%
\section{Soft Constraint Optimization}%

Soft constraints are constraints which should be respected if possible.  When not all soft constraints can hold simultaneously we attach a \emph{cost} to each violation.  In the resulting optimization problem the overall cost is to be minimized.  Soft constraints may be intensional or extensional.  An intensional constraint is an equation or predicate capturing the desired relationship between variables, whereas an extensional constraint is written as a table with columns for the variables of interest, explicitly listing the allowed or disallowed tuples.

Specialized solvers have been highly successful for soft-constraint problems in extensional form.  All of these solvers attempt to discover conflicts between soft constraints, or unsatisfiable soft constraints, by using what are essentially lookahead approaches, followed by appropriate reformulation that exposes the increased lower bound on solution cost due to the conflict or violation.

For WCSP, in which each extensional table contains a weight column giving the cost to be paid if the row holds, the best solver seems to be \toolbar{}/\toulbar{}~\cite{degivry,larrosa}.  It is based on a branch-and-bound search with consistency notions, where (loosely speaking) a variable is consistent if the \emph{costs of the minimum cost value(s)} have been subtracted from the tables involving the variable and moved into the lower bound, thus \emph{fathoming} unpromising branches.

For MAXSAT, good solvers in a recent evaluation~\cite{maxsat} included \akmaxsat{}~\cite{kuegel} and \maxsatz~\cite{lin} variants.  Essentially they use lookahead, with unit propagation and failed literal detection, to improve the lower bound~\cite{li2,li}.  In restricted cases they use MAXSAT resolution, in which conflicting clauses are replaced by a unified clause plus compensation clauses~\cite{bonet}.  WCSP solvers are also highly effective on MAXSAT, since MAXSAT is a special case of WCSP.

Recently there has also been considerable interest in decomposing MAXSAT to SAT, usually with an unsatisfiable-core approach~\cite{ansotegui,fu,marquessilva,marquessilva3}.  Because they use learning instead of lookahead (and other improvements such as activity-based search~\cite{moskewicz}), they have a considerable advantage over the previously-described approaches.  On the other hand they do not employ reformulation, and not all problems are suitable for unsatisfiable-core searches.

Pseudo-Boolean Optimization (PBO) is also promising for MAXSAT, which is a special case of PBO.  In particular the Weighted Boolean Optimization (WBO) framework~\cite{manquinho} is an application of PBO to soft-constraint problems, using some of the specialized techniques discussed above.  Another option is decomposition to SAT via the PBO solver MiniSAT+~\cite{een}, which could be useful if unsatisfiability-based methods aren't applicable to a particular problem.

In this research we extend certain of the above techniques to \emph{intensional} soft constraints.  Modelling with intensional constraints has many advantages, \begin{inparaenum}[(i)]\item it is much easier since constraints are expressed in a natural way, \item it handles more constraints, since decomposition to extensional form is not always practical, and \item it can be more efficient, since propagation is a reasoning task as opposed to an expensive table traversal\end{inparaenum}.  It also has some disadvantages, \begin{inparaenum}[(i)]\item propagators must be implemented for each type of intensional constraint, and \item due to the many ways that constraints can interact, reformulation is difficult\end{inparaenum}.

We consider solving 
combinatorial constrained optimization problems
with \emph{pseudo-Boolean objective} (COPPBO).
A COPPBO $(\mathbf{x}, \mathbf{y}, \mathbf{D}, \mathbf{C}, \mathbf{c})$
consists of a vector $\mathbf{x}$ of general variables $x_i$, $i \in 1..m$,
a vector of $\mathbf{y}$ of Boolean variables $y_i$, $i \in 1..n$ which
appear in the objective, 
an initial finite domain $\mathbf{D} = D(x_1) \times \ldots \times D(x_m)
\times D(y_1) \times \ldots \times D(y_m)$, 
a set $\mathbf{C}$ of constraints $C_i$, $i \in 1..k$ 
and an objective $z = \mathbf{c}^T \mathbf{y}$ to be minimized, 
where $\mathbf{c}$ consists of positive constant weighting factors\footnote{
In calculating $\mathbf{c}^T \mathbf{y}$ we take 
$\false = 0$ and $\true = 1$.}\footnote{We make the coefficients $c_i$ positive
by negating Boolean literals if necessary.}.  COPPBO problems encompass
MAXSAT, partial MAXSAT, weighted partial MAXSAT, WBO and PBO.

An important class of COPPBO problems are
\emph{soft constraint optimization problems}
$(\mathbf{x}, \mathbf{D}, \mathbf{H}, \mathbf{S}, \mathbf{c})$ given
by a set of hard constraints $\mathbf{H}$ and a vector of soft constraints
$\mathbf{S}$, with corresponding weight vector $\mathbf{c}$
such that $c_i$ is the cost of violating soft
constraint $S_i$, $i \in 1..n$.  The aim is to find a solution
$\theta \in \mathbf{D}$ to the variables $[\mathbf{x}\:\mathbf{y}]$,
which minimizes $z = \mathbf{c}^T \mathbf{y}$,
subject to $\mathbf{H} \cup \{\neg y_i \rightarrow S_i: i \in 1..n\}$,
where $\mathbf{y}$ consists of introduced \emph{relaxation variables} for the
constraints in $\mathbf{S}$.
Note that a CP system supporting constraint $S_i$ can be
straightforwardly
extended to support the softened form $\neg y_i \rightarrow S_i$ through
half-reification~\cite{feydy}.  

\section{Basic Unsatisfiable Cores Algorithm}\label{sec:basic}%

An \emph{unsatisfiable core} is a clause which contains only literals in
$\mathbf{y}$.  This clause forces an objective variable to be $\true$, and must
add some positive value to the objective.
Note that clauses containing a literal $\neg y_i$ are not unsatisfiable cores.
The unsatisfiable cores approach to optimization originally arose for
solving a MAXSAT problem, which is, given a set of soft clauses $S$,
find a solution which satisfies the maximum number of soft clauses.

\begin{algorithm}%
\begin{algorithmic}[1]%
\Require{$F = \{y_1, \ldots, y_n\}$ initially}
\Function{Learn}{$\mathbf{C}, L, M$} where $\abs{M} > 1$ (otherwise fall back to Algorithm~\ref{alg:dpll})
  \State $F \leftarrow F - M$\tabto{8em}\Comment{remove candidates that have appeared in an unsatisfiable core}
  \State \Return{$\mathbf{C}$}\tabto{8em}\Comment{unchanged constraint set, to avoid risk of learning duplicates}
\EndFunction
\vspace*{2em plus 3em minus 1em}%
\Require{$F =$ set of $y_i$ literals which have never appeared in an unsatisfiable core}
\Function{Decide}{$D(x_1) \times \ldots \times D(x_m) \times D(y_1) \times \ldots \times D(y_n)$}
  \If{$\textit{decision\_level = 0}$ and exists $y_i \in F$ with $D(y_i) = \{\false, \true\}$}
    \State for all such $y_i$ restrict the corresponding domain $D(y_i)$ to $\{\false\}$
  \Else
    \State restrict some other domain according to user's programmed search
  \EndIf
  \State \Return{$D(x_1) \times \ldots \times D(x_m) \times D(y_1) \times \ldots \times D(y_n)$}
\EndFunction
\end{algorithmic}
\caption{Basic unsatisfiable core algorithm (relative to Algorithm~\ref{alg:dpll})}\label{alg:basic}
\end{algorithm}
The basic unsatisfiable core solver consists of the procedures in
Algorithm~\ref{alg:basic}, which are called by the high level solver of
Algorithm~\ref{alg:dpll}, and essentially modify the decision procedure
based on information from conflict analysis.  Each attempt fixes all
\emph{unfixed variables in $\mathbf{y}$, that have never appeared in an
  unsatisfiable core}, to $\false$, and solves the resulting problem.  
This either finds a solution (which should be of low cost), or it detects
that the problem is unsatisfiable.  
In a learning solver such as a SAT or LCG solver, 
by fixing the $\mathbf{y}$-variables as a 
(possibly) \emph{multiple decision} in an artificial first decision level, 
failure occurring at this level generates, as a side effect, 
a new unsatisfiable core.  This continues until solutions are found, or the original problem is proved unsatisfiable.

We have to modify the standard LCG solver to 
allow multiple simultaneous decisions
when branching in procedure \textproc{Decide}. The
\textproc{Analyze} procedure returns \emph{generalized 1UIP nogoods} $L
\rightarrow M$ where $L$ is a set of literals treated as a conjunction,
and $M$ a set treated as a disjunction, e.g. $a \wedge b \rightarrow c \vee
d$.  The code for \textproc{Analyze} in Algorithm~\ref{alg:dpll} already 
handles this case, as line~\ref{lin:exit} will
cause the loop to exit when only decisions remain.  
This line is unnecessary
for search without multiple simultaneous decisions.
Note also that, for the basic unsatisfiable core algorithm, we will only
ever generate generalized 1UIP nogoods where $L = \emptyset$, but
we will use the more general form in the next section.
 
Algorithm~\ref{alg:basic} keeps track of $F$, the set of optimization
variables that have never appeared in an unsatisfiable core.
The new post-analysis handler $\textproc{Learn}(\mathbf{C}, L, M)$ 
handles the case where $\abs{M} > 1$, by removing variables in $M$ 
from $F$.  Unlike the 1UIP case with $\abs{M} = 1$, it does not learn the new
nogood $L \rightarrow M$ since this may already be in the clause database,
because two or more literals in $M$ may have been set $\false$ simultaneously
(propagating the database precludes single wrong decisions that
violate a clause, but this does not extend to multiple decisions).

\begin{figure}[t]
\centering
\begin{tabular}{c@{\hspace{2cm}}c}
$
\xymatrix@=2mm{
& ~\ar@{--}[ddddd] \\
\neg y_1 \ar[rr] && \neg b \\
\neg y_2 \\
\neg y_3 \ar[rr] && \neg a \ar[r] & \false \\
\neg y_4 \ar[rrru] \\
& ~
}
$
&
$
\xymatrix@=2mm{
& ~\ar@{--}[ddddd] \\
\neg y_1 \ar[rr] && \neg a \ar[ddr] \\
\neg y_2 \ar[rr] && \neg b \ar[dr] \\
\neg y_3 \ar[rrr] &&& \false \\
\neg y_4 \ar[rr] && \neg c \\
& ~
}
$
\\
(a) & (b)
\end{tabular}
\caption{Two implication graphs from solving the soft constraint
  problems of: (a) Example~\ref{ex:maxsat}, and (b) 
Example~\ref{ex:branchandbound}; assuming all soft
 constraints are satisfied.\label{fig:maxsat}}
\end{figure}
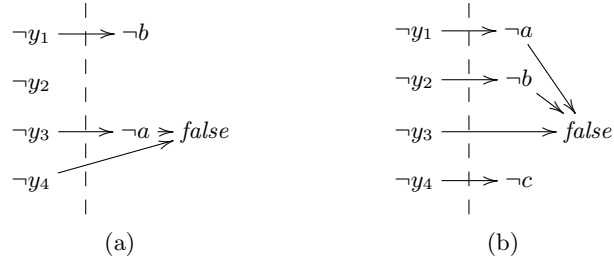

\begin{example}\label{ex:maxsat}%
The MAXSAT instance (all constraints soft) 
in the left column is entered as on the right,
with objective $z = y_1 + y_2 + y_3 + y_4$ to be minimized.
\ignore{
\begin{align*}%
C_1 &\equiv \neg b & &y_1 \vee \neg b \\
C_2 &\equiv a \vee b & &y_2 \vee a \vee b \\
C_3 &\equiv \neg a & &y_3 \vee \neg a \\
C_4 &\equiv a & &y_4 \vee a.
\end{align*}%
}
\\
\centerline{
$
\begin{array}{rclcrcl@{~~~}|@{~~~}lcl}%
C_1 &\equiv & \neg b &~~~& C_2 &\equiv& a \vee b &y_1 \vee \neg b &~~~&y_2
\vee a \vee b\\
C_3 &\equiv & \neg a  &&C_4 &\equiv& a & y_3 \vee \neg a && y_4 \vee a \\
\end{array}
$}
Here $C_3$ and $C_4$ are clearly in conflict, but given the choice it is
better to relax $C_3$ so that $C_1$ and $C_2$ can be satisfied
simultaneously.  
At the top level the solver creates the multiple decision
(simultaneously) $y_1 = \false$, $\ldots$, $y_4 = \false$ 
and solves.
This fails with unsatisfiable core $y_3 \vee y_4$, which is
the generalized 1UIP nogood resulting from
the implication graph in Figure~\ref{fig:maxsat}(a).

On the next attempt it sets only $y_1 = \false$, $y_2 = \false$, 
because $y_3$, $y_4$ have appeared in an unsatisfiable core.  
The only possible solution, under these assumptions, is
$b = \false$, $a = \true$, $y_3 = \true$, $y_4 = \false$ with 
cost $1$.
\end{example}%

Although the initial solutions found by the unsatisfiable core
algorithm are frequently of low cost, the algorithm does not replace the
need for a traditional branch-and-bound approach (or equivalently the
solving of a series of satisfaction problems with tighter and tighter bounds
on the cost).  When we have a (best known) solution $\theta$, the objective
bound $\mathbf{c}^T \mathbf{y} < \mathbf{c}^T \theta(\mathbf{y})$ 
needs to be propagated.
This is standard for CP systems, although  
MAXSAT solvers typically use SAT decompositions of the objective
using BDDs or sorting networks~\cite{marquessilva}.

\begin{example}\label{ex:branchandbound}%
Consider the trivial instance (original on left, reified on right),
\ignore{
\begin{align*}%
C_1 &\equiv \neg a & &y_1 \vee \neg a \\
C_2 &\equiv \neg b & &y_2 \vee \neg b \\
C_3 &\equiv a \vee b & &y_3 \vee a \vee b \\
C_4 &\equiv \neg c & &y_4 \vee \neg c.
\end{align*}%
}
\\
\centerline{
$
\begin{array}{rclcrcl@{~~~}|@{~~~}lcl}%
C_1 &\equiv & \neg a &~~~& C_2 &\equiv& \neg b &y_1 \vee \neg a &~~~&y_2 \vee \neg b\\
C_3 &\equiv & a \vee b &&C_4 &\equiv& \neg c &y_3 \vee a \vee b && y_4 \vee \neg c\\
\end{array}
$}
The only unsatisfiable core is $y_1 \vee y_2 \vee y_3$, discovered from the
implication graph in Figure~\ref{fig:maxsat}(b).
The solver will never try $c = \true$ because 
$y_4$ does not participate in
any unsatisfiable core.  
After discovering the core, search will set $y_4 = \false$
and propagate $\neg c$. Search will continue
finding solutions with $a$ = $b = \true$ at cost 2 or any other
valuation to $a$, $b$ at cost 1.  
Unsatisfiable-core algorithms do not distinguish between these cases.
\end{example}%
Superficially the algorithm is a static search which fixes the variables in
$\mathbf{y}$ to $\false$ before proceeding, but is quite different because 
\begin{inparaenum}[(i)]
\item any propagation resulting
  from fixing the $\mathbf{y}$-variables is done after \emph{all} are fixed,
  which is enormously more efficient than one-by-one, especially when global
  constraints are involved, and 
\item static search backtracks only as
  far as necessary, so it will try all values 
for the tuple $(y_1,
  \ldots, y_n)$ in lexical order, which is impractical for large
  $n$\end{inparaenum}.

In the best MAXSAT implementations~\cite{marquessilva,marquessilva3} the
relaxation
variables for a clause were not created until the clause had appeared in an
unsatisfiable subset, with the information about conflicting clauses being
extracted from proof traces rather than from the presence of their
relaxation literals in a learnt clause.  This was sensible for MAXSAT
considering that MAXSAT problems are `tall', i.e.~they have more clauses
than variables, hence more relaxation variables than ordinary variables,
unless relaxation variables are created lazily.

On the other hand CP has much less reliance on decomposition due to its
richer constraint library and problems are frequently `wide', i.e.~they have
more variables than constraints, particularly when there is extensive use of
global constraints.  So relatively few relaxation literals are required and
it was not worth going to the trouble of reformulating the problem `on the
fly', so we opted for the simpler approach of the artificial decision level,
which allows conflict analysis to collect relaxation literals into
unsatisfiable cores in a natural way.

\section{Nested Unsatisfiable Core Algorithm}\label{sec:nested}%

With the previously outlined approach to unsatisfiable cores it can happen that all or most variables in $\mathbf{y}$ are covered by one or more unsatisfiable cores.  This is particularly likely in problems with a lot of structure or symmetry such that each variable is involved in a similar set of constraints.  In such cases, once all unsatisfiable cores have been enumerated, all soft constraints have reverted to being treated as soft, and the problem is no easier than originally.

Hence we define a nested version of the algorithm in which soft constraints
can be made hard during search rather than only at the top level.  This
involves two new concepts, \begin{inparaenum}[(i)]\item \emph{active}
  unsatisfiable cores which describe a conflicting set of soft constraints
  where we do not know which will be violated, as opposed to \emph{inactive}
  where a particular violation is known and has already been
  penalized, and \item \emph{contingent} unsatisfiable cores which are only
  unsatisfiable cores in the current subproblem, that is, under the current
  search assumptions\end{inparaenum}.

Note that in the clausal view a \emph{contingent} unsatisfiable core is simply any clause where only $y_i$ literals remain, because other literals are already $\false$.  Then at least one of these $y_i$ must be true, i.e.~it is an unsatisfiable core.  It is initially \emph{active} but goes \emph{inactive} when at least one of these $y_i$ becomes $\true$.

In the revised algorithm, an objective variable $y_i$ may be set $\false$
initially by a multiple decision,
but then reverts to unconstrained when search discovers it is part of an
unsatisfiable core and backtracks.  
It will no longer be set to $\false$ since it appears in an 
active unsatisfiable core. But 
it may again be set $\false$ at a deeper decision level if all 
the unsatisfiable cores involving $y_i$ go inactive. 
\ignore{
If solving then backtracks to the decision level
where it was set $\false$, with a contingent unsatisfiable core again
involving $y_i$, it is made unconstrained (and will not be set
$\false$ again until any new unsatisfiable core involving it becomes
inactive).
}

\begin{example}%
Referring to Example~\ref{ex:branchandbound}, given the unsatisfiable core
$y_1 \vee y_2 \vee y_3$, on the next attempt $y_4$ has not been involved in
any core and so is set to $\false$ on level 1, hence $c = \false$.  
On level 2, suppose 
search tries $a = \true$ hence $y_1 = \true$ and reaches a fixed point.  
Since $y_1 \vee y_2 \vee y_3$ is now satisfied, there is 
no evidence that $y_2$ or $y_3$ needs to be $\true$, and a 
new multiple decision is made as level 3 which sets $y_2 = y_3 = \false$.  
Solving continues, finding $b = \false$ and cost 1.
\end{example}%
\begin{example}\label{ex:nesting}%
Consider minimizing $a + b + c$ subject to the clauses
\[
a \vee b \vee d  \hspace*{2cm}
b \vee c \vee \neg d \hspace*{2cm}
\neg a \vee d,
\]
\ignore{
\begin{align*}
&a \vee b \vee d \\
&b \vee c \vee \neg d \\
&\neg a \vee d,
\end{align*}
}
which arises from the partial MAXSAT instance with soft constraints $\neg a, \neg b, \neg c$ and hard constraints as above.  Falsifying $a$, $b$, $c$ returns the unsatisfiable core $a \vee b \vee c$, so that on the next attempt, nothing can be fixed at the top level.  This is exactly the problem that the nested algorithm tries to address.

Since a multiple decision is not applicable the ordinary decision procedure sets $a = \true$ on level 1, resulting in $d = \true$, and reaches a fixed point.  Because $a \vee b \vee c$ is now satisfied, level 2 is a multiple decision setting $b = c = \false$, but this immediately fails with reason $b \vee c \vee \neg d$, because $d$ was already $\true$.  Solving backtracks past the multiple decision and records the new unsatisfiable core $b \vee c$ (contingent on $d = \true$), then continues, using the ordinary decision procedure, eventually finding a solution of cost 2 (with $a$ and either $b$ or $c = \true$).
\end{example}%
The benefit of this approach is that even if all objective literals
eventually appear in an unsatisfiable core, we still aggressively set them to
$\false$ as soon as all the active (contingent) unsatisfiable cores they
appear in are satisfied.  Conversely, it may be there are too many
contingent unsatisfiable cores to be enumerated by this method, leading to
thrashing in the solver.

\begin{algorithm}[t]
\begin{algorithmic}[1]%
\Require{$a_i =$ count of \emph{active} contingent unsatisfiable cores involving $y_i$, initially 0}
\Require{$Y_i =$ stack of \emph{all} contingent unsatisfiable core involving $y_i$, initially empty}
\vspace*{2em plus 3em minus 1em}%
\Require{When literal $y_i$ goes $\true$ due to propagation or decision,}
\For{$\{L \rightarrow M\}$ in stack $Y_i$ where $\{L \rightarrow M\}$ is still active}
  \State mark $\{L \rightarrow M\}$ as inactive, and for all $y_j \in M$ decrement $a_j$
\EndFor
\vspace*{2em plus 3em minus 1em}%
\Function{Learn}{$\mathbf{C}, L, M$} where $\abs{M} > 1$ (otherwise fall back to Algorithm~\ref{alg:dpll})
  \State for all $y_i \in M$ push the clause $\{L \rightarrow M\}$ onto stack $Y_i$ and increment $a_i$
  \State \Return{$\mathbf{C}$}\tabto{8em}\Comment{unchanged constraint set, to avoid risk of learning duplicates}
\EndFunction
\vspace*{2em plus 3em minus 1em}%
\Function{Decide}{$D(x_1) \times \ldots \times D(x_m) \times D(y_1) \times \ldots \times D(y_n)$}
  \If{exists $y_i$ with $a_i = 0$ and $D(y_i) = \{\false, \true\}$}
    \State for all such $y_i$ restrict the corresponding domain $D(y_i)$ to $\{\false\}$
  \Else
    \State restrict some other domain according to user's programmed search
  \EndIf
  \State \Return{$D(x_1) \times \ldots \times D(x_m) \times D(y_1) \times \ldots \times D(y_n)$}
\EndFunction
\end{algorithmic}
\caption{Nested unsatisfiable core algorithm (relative to Algorithm~\ref{alg:dpll})}\label{alg:nested}
\end{algorithm}
The revised solver replaces Algorithm~\ref{alg:basic} with
Algorithm~\ref{alg:nested} which maintains activity counts $a_i$ for each
variable $y_i$.
When a new core becomes active through discovery in conflict analysis,
the counts for all variables involved 
are incremented. When the core goes inactive, the counts are decremented,
based on a cross-referencing data structure consisting of stacks $Y_i$.
The counts $a_i$ and stacks $Y_i$ must be trailed so that they are reset
correctly on backjumping.  The \textproc{Decide} procedure only
considers $y_i$ variables with zero activity counts as candidates for
a multiple decision, or if none exist it makes a normal search decision.

\section{Notification-based Nested Algorithm}\label{sec:notification}%

In the algorithms discussed to this point, we did not `learn'
unsatisfiable cores by adding them to the constraint store, because we did
not want to risk learning duplicates, leading to duplicated
propagation work.
This is frustrating for the nested algorithm
since contingent unsatisfiable cores are discarded on backtracking,
even though they are valid nogoods which may capture new information.

\begin{example}%
In Example~\ref{ex:nesting}, the first unsatisfiable 
core is $a \vee b \vee c$ 
which holds new information not explicit in the original 
formulation, that can help future search. 
The next (contingent) unsatisfiable core is $b \vee c \vee \neg d$ 
which is a duplicate of a clause in the original problem.  Other than
searching the constraint store or using a hashing scheme, 
we have no way of distinguishing between these cases.
\end{example}%
Furthermore, with the previously-defined algorithms, 
clausal constraints from the constraint store are not used
as contingent unsatisfiable cores except
when they are violated by a multiple decision.  
To tackle this problem we modify the clausal propagators to issue a notification whenever an ordinary clause becomes an active contingent unsatisfiable core, that is, when the last of the non-$y_i$ literals in the clause goes $\false$.

In essence we treat the clause $C = X \vee Y$, where $Y$ are the $y_i$
literals in $C$ and $X$ the others, as $X \vee l$ and $\neg l \vee Y$,
where $l$ is a new literal.
When $l$ is set $\true$ by unit propagation, the 
clause $C$ becomes a contingent unsatisfiable core.  This is preferable 
because any multiple decisions made by the unsatisfiable core algorithms 
take into account all current information 
and do not lead to trivial failures.  
Then it is safe to learn all clauses 
resulting from conflict analysis.

\ignore{
\pjs{This example is not clear, since it has the reformulation step. omit it}
\begin{example}%
Revisiting Example~\ref{ex:branchandbound}, it is sensible to reformulate the problem as minimizing $a + b + y_1 + c$ subject to the single clause $y_1 \vee a \vee b$.  The notification-based solver realizes during initialization that this is an unsatisfiable core, skips the first failed solving attempt, and finds the first solution immediately.
\end{example}%
}%
\begin{example}%
Revisiting Example~\ref{ex:nesting}, the first unsatisfiable core $a \vee b
\vee c$ has to be derived in the usual way by conflict analysis resulting
from a multiple decision, but on the second attempt, as
soon as $d$ is set $\true$ by propagation from $a$, the \emph{original} 
clause $b \vee c \vee \neg d$ becomes an active 
contingent unsatisfiable core and is immediately recorded against the literals $b$ and $c$ as evidence that one of them must be $\true$.  This avoids the failed search with $b = c = \false$.
\end{example}%
\section{Enhanced Lower Bounding}\label{sec:enhanced}%

At each node of the branch and bound tree we need to check the \emph{linear}
constraint $\mathbf{c}^T \mathbf{y} < \mathbf{c}^T \theta(\mathbf{y})$ 
where $\theta$ is the current best solution and fathom (backtrack) when the
constraint detects failure.  This compares the current solution cost, or if
we only have a partial solution then a conservative \emph{lower bound} on
the solution cost, with a conservative upper bound on the \emph{optimal}
solution cost.  Usually the conservative lower bound is obtained by taking all
unfixed $y_i$ as $\false$.

Information from unsatisfiable cores can be used to strengthen the lower bound.  Suppose that some subset e.g.~$y_2$, $y_3$ and $y_5$ are unfixed but we have the unsatisfiable core $y_2 \vee y_3 \vee y_5$.  Then clearly it is too conservative to assume that all are $\false$ and we can increase the estimate by at least $\min(c_2, c_3, c_5)$, hopefully leading to earlier fathoming.  This is the basis of \emph{disjoint inconsistent subformula} approaches which have been used for MAXSAT~\cite{li}.

\begin{example}%
Referring to Example~\ref{ex:branchandbound}, after a finding the first solution of cost 1, the unsatisfiable core $y_1 \vee y_2 \vee y_3$ implies that any solution will have cost $\ge 1$ so no further improvement is possible.  Solving can then terminate immediately.
\end{example}%

Note that our treatment is somewhat more formal as LCG solvers require
\emph{reasons} for fathoming.  Thus it is inadequate just to increase the
estimate, we have to derive a globally true but tighter objective
constraint.  Suppose we have an objective constraint $L \equiv \sum_{i =
  1}^n c_i y_i < u$ where $u$ is the current upper bound, and a collection
of \emph{active} unsatisfiable cores $C_1$, $\ldots$, $C_\ell$.  Then
considering core $C_i$ as a \emph{linear} constraint
$G_i \equiv \sum_{l \in C_i} -l \le -1$, we can produce a tightened objective
constraint as a linear combination of $L$ and $G_i$, $i \in 1..\ell$.

\begin{example}\label{ex:disj}
Suppose the problem is minimizing $2y_1 + 3y_2 + 3y_3 + 5y_4$ with $u = 7$ 
so that the objective constraint is $U \equiv 2y_1 + 3y_2 + 3 y_3 + 5 y_4 < 7$.
Unsatisfiable core $C_1 \equiv y_1 \vee y_3 \vee y_4$ gives the
\emph{linear} constraint $G_1 \equiv -y_1 - y_3 - y_5 \leq -1$.  Then a
strengthened objective constraint is obtained as the Fourier
elimination~\cite{fourier}  of
$y_1$, by summing the constraints $U' \equiv U + 2 G_1 
\equiv 3y_2 + y_3 + 3y_5 < 5$. 
Unsatisfiable core $C_2 \equiv y_2 \vee y_3 \vee y_4$ 
gives
$G_2 = -y_2 - y_3 - y_4 \leq 1$ 
and eliminates $y_3$ giving $U'' \equiv U' + G_2 \equiv 2y_2 + 2y_5 < 4$,
which is tighter while still globally true.
\end{example}%
We use two methods to strengthen the objective constraint $U$,
\begin{inparaenum}[(i)]\item a heuristic method and
\item an exact method\end{inparaenum}.
Note that the tightened inequality is contingent upon the non-$y_i$ literals 
of any contingent unsatisfiable cores that contributed to tightening, 
so whenever we have to explain the actions of the tightened propagator $U$
we have to add these literals to the explanation.

\subsubsection{Disjoint Unsatisfiable Core-based Lower Bounding:}\label{sec:disjoint}%

\begin{algorithm}[t]
\begin{algorithmic}
\Function{Disjoint}{$\mathbf{c}^T \mathbf{y} < u, \{G_1, \ldots, G_\ell\}$}
  \For{$i$ in $1..\ell$}
    \State $\alpha_i \leftarrow \min_{y_j \in G_i} c_j$
    \State \algorithmicforall{} $y_j$ in $G_i$ \algorithmicdo{} $c_j
    \leftarrow c_j - \alpha_i$ \algorithmicend{} \algorithmicfor{}
    \State $u \leftarrow u - \alpha_i$ 
  \EndFor
  \State \Return $\mathbf{c}^T \mathbf{y} < u$
\EndFunction
\end{algorithmic}
\caption{Disjoint unsatisfiable core-based bound-strengthening}\label{alg:disjoint}
\end{algorithm}
The heuristic method, shown as Algorithm~\ref{alg:disjoint}, starts by
taking a working version of the objective upper-bound constraint $U$.  Then
for each $G_i$ it eliminates from $U$, by the Fourier-Motzkin
method~\cite{fourier}, the cheapest variable that could be made $\true$ to
satisfy $G_i$, using the current working coefficients as costs.  This means
adding the assumed cost of satisfying $G_i$ into the estimate, then
adjusting the assumed costs used from then on, to avoid double-counting.  
Example~\ref{ex:disj} illustrates the algorithm.

The above algorithm is well known for weighted MAXSAT.  We define an
incremental version that modifies the objective constraint each time new
active unsatisfiable cores are discovered.  When unsatisfiable cores go
inactive they are removed from the tightened inequality, 
which is necessary for correctness, 
since the tightened bound constraint 
is a \emph{consequence} of the original bound constraint, not equivalent.  After removing an inactive core, we found it essential to re-examine subsequently-added cores to see if they can contribute any additional strength.

\begin{example} 
Continuing Example~\ref{ex:disj}, suppose later $y_1 = \true$,
then $U''$ does not propagate, but $U$ would have since it then becomes
$3y_2 + 3y_3 + 5 y_4 < 5$ which immediately sets $y_4 = \false$. 
We roll back to $U$
where we eliminated the first core involving $y_1$, and
now eliminate the remaining active core $C_2 \equiv y_2 \vee y_3
\vee y_4$  to obtain 
$U''' \equiv U + 3 G_2 \equiv 2y_1 + 2 y_4 < 4$, which propagates $y_4 =
\false$.
\end{example}

\subsubsection{Linear Programming-based Lower Bounding:}\label{sec:lp}%

The strengthening procedure is an optimization problem over the coefficients $\alpha_i$ discussed above, which can be solved to optimality using a Linear Programming (LP) solver to give the best possible fathoming based on the information available at each node.  The LP has to be updated at each node with the latest set of unsatisfiable cores, noting that inactive cores can be left in, since it can see they are already satisfied.  The LP can also contain other constraints at the modeller's discretion.  

We define an LP constraint $\textit{linear\_program}(\mathbf{G}, \mathbf{c}, \mathbf{y}, z)$ which given a set of linear (in)equalities $\mathbf{G}$ over the variables $\mathbf{y}$, an objective $\mathbf{c}^T \mathbf{y}$ and an upper bound $z$ on the objective (based on the best solution found so far), enforces that all (in)equalities are satisfied and $\mathbf{c}^T \mathbf{y} \le z$.  It executes when the bounds on $\mathbf{y}$ are tightened, and verifies that $\mathbf{G}$ can still hold, if not it detects failure with an explanation.  It also verifies that there is still objective slack, if not it fathoms with an explanation.  Then it prunes the $\mathbf{y}$ if possible, based on the slack.

The \textit{linear\_program} propagator works similarly to the bound-strengthening procedure described previously, in that it derives new \linear{} constraints $L + \sum_{i \in 1..\ell} \alpha_i G_i$, choosing the vector $\alpha$ which minimizes the RHS of the resulting constraint.  For the LP case the vector $\alpha$ and indeed the coefficients and RHS of the strengthened constraint are available directly in the dual solution after minimizing $\mathbf{c}^T \mathbf{y}$ subject to $\mathbf{G}$.  Thus the standard procedure of deriving explanations from dual solutions or unbounded dual rays is applicable~\cite{achterberg,davey,downing2}.

\section{Experiments}

\ignore{
Fortunately all of the important primitive constraints and a number of soft globals are already available for LCG solvers~\cite{ohrimenko,downing,downing2}, but we aren't aware of a comprehensive soft-constraint problem library (disregarding Max/SoftCSP~\cite{maxcsp,softcsp} and MAXSAT~\cite{maxsat} libraries as not having rich enough constraints).  The Minizinc 1.4.2 benchmarks library had some CP-optimization problems which provided a basis~\cite{minizinc} but we also extended the set with extra \minizinc{} models created for this paper, based on standard problems from the literature~\cite{nsplib,ortega,decesco}\nd{revisit}.
}%
To evaluate whether unsatisfiable cores are useful for CP and whether our
extensions are helping, we compared standard branch-and-bound with the
solvers described in Sections \ref{sec:basic}, \ref{sec:nested} and
\ref{sec:notification}.  With each of these solvers we tried standard lower
bounding and each of the strengthened lower bounds from
Section~\ref{sec:enhanced}.

The basic solver is a state-of-the-art LCG solver, \chuffed{}.
We use activity-based search (VSIDS) for all
experiments~\cite{moskewicz}.  For maximization problems we negate the
objective, so all are minimization.  The experiments were run on a cluster
of AMD 6-core Opteron 4184 at 2.8 GHz with time limit 3600s,
memory limit 2 Gbytes per core, and memory-outs treated as timeouts.
Data files are available from
\url{http://www.csse.unimelb.edu.au/~pjs/unsat_core}.

\begin{table}[t]%
\caption{Evaluating unsatisfiable-core solvers on combinatorial benchmarks}\label{tab:bounding}%
{\centering \scriptsize \begin{tabular}{|l@{\ }r|r@{}r@{}r@{}r|r@{}r@{}r@{}r|r@{}r@{}r@{}r|}
\hline
\multicolumn{2}{|l|}{\T} & \multicolumn{4}{l|}{fathoming=std} & \multicolumn{4}{l|}{disjoint} & \multicolumn{4}{l|}{empty LP} \\
\multicolumn{2}{|l|}{} & \multicolumn{4}{r|}{opt,s,sol,obj} & \multicolumn{4}{r|}{opt,s,sol,obj} & \multicolumn{4}{r|}{opt,s,sol,obj} \\
\hline
\T \model{psm} (k=1) & branch and bound & 12, & 1117s, & 16, & -672 & 12, & 1015s, & 16, & -674 & 11, & 1214s, & 16, & -678 \\
& basic unsat core & 12, & 1084s, & 16, & -642 & {\bf 12}, & {\bf 966s}, & {\bf 16}, & {\bf -643} & 11, & 1285s, & 16, & -641 \\
& nested unsat core & 8, & 1943s, & 16, & -629 & 12, & 1461s, & 16, & -643 & 7, & 2353s, & 16, & -599 \\
& nested+notification & 10, & 1658s, & 16, & -644 & 11, & 1551s, & 16, & -648 & 12, & 1341s, & 16, & -645 \\
\hline
\T \model{psm} (k=2) & branch and bound & 2, & 3244s, & 16, & -654 & 1, & 3375s, & 14, & -675 & 1, & 3375s, & 16, & -724 \\
& basic unsat core & 0, & 3600s, & 13, & -581 & 1, & 3375s, & 13, & -574 & 2, & 3248s, & 14, & -567 \\
& nested unsat core & 0, & 3600s, & 10, & -559 & 2, & 3161s, & 13, & -579 & 2, & 3160s, & 13, & -596 \\
& nested+notification & 2, & 3167s, & 16, & -640 & 2, & 3157s, & 16, & -626 & {\bf 4}, & {\bf 2973s}, & {\bf 16}, & {\bf -602} \\
\hline
\T \model{photo} & branch and bound & 10, & 2487s, & 30, & -15.7 & 10, & 2492s, & 30, & -15.5 & 10, & 2460s, & 30, & -15.6 \\
& basic unsat core & 11, & 2455s, & 30, & -15.9 & 14, & 2211s, & 30, & -15.7 & 14, & 2083s, & 30, & -15.8 \\
& nested unsat core & 10, & 2499s, & 30, & -15.8 & 17, & 1781s, & 30, & -15.9 & 21, & 1268s, & 30, & -15.9 \\
& nested+notification & 12, & 2437s, & 30, & -15.8 & 18, & 1609s, & 30, & -15.9 & {\bf 26}, & {\bf 926s}, & {\bf 30}, & {\bf -16.0} \\
\hline
\T \model{rlfap} & branch and bound & 1, & 3380s, & 16, & 11.0 & 1, & 3379s, & 16, & 10.8 & 1, & 3379s, & 16, & 10.8 \\
& basic unsat core & 2, & 3154s, & 16, & 9.7 & {\bf 3}, & {\bf 3086s}, & {\bf 16}, & {\bf 9.5} & 2, & 3154s, & 16, & 9.7 \\
& nested unsat core & 2, & 3156s, & 14, & 9.3 & 2, & 3170s, & 15, & 9.1 & 1, & 3375s, & 14, & 9.4 \\
& nested+notification & 2, & 3154s, & 14, & 9.2 & 2, & 3174s, & 15, & 9.1 & 2, & 3175s, & 14, & 9.2 \\
\hline
\T \model{roster} & branch and bound & 14, & 1316s, & 20, & 15.7 & 14, & 1297s, & 20, & 15.6 & 14, & 1197s, & 20, & 16.4 \\
& basic unsat core & 16, & 724s, & 16, & 15.4 & 16, & 724s, & 16, & 15.4 & 16, & 724s, & 16, & 15.4 \\
& nested unsat core & 16, & 724s, & 16, & 15.4 & 16, & 724s, & 16, & 15.4 & 16, & 724s, & 16, & 15.4 \\
& nested+notification & {\bf 16}, & {\bf 723s}, & {\bf 16}, & {\bf 15.4} & 16, & 725s, & 16, & 15.4 & 16, & 724s, & 16, & 15.4 \\
\hline
\T \model{sugiyama} & branch and bound & 5, & 99s, & 5, & 8.6 & 5, & 106s, & 5, & 8.6 & 5, & 121s, & 5, & 8.6 \\
& basic unsat core & 5, & 166s, & 5, & 8.6 & 5, & 143s, & 5, & 8.6 & 5, & 102s, & 5, & 8.6 \\
& nested unsat core & 5, & 128s, & 5, & 8.6 & 5, & 5s, & 5, & 8.6 & 5, & 63s, & 5, & 8.6 \\
& nested+notification & 5, & 207s, & 5, & 8.6 & 5, & 14s, & 5, & 8.6 & {\bf 5}, & {\bf 3s}, & {\bf 5}, & {\bf 8.6} \\
\hline
\end{tabular}
}%
\end{table}%
We tried the following combinatorial benchmarks:  
\model{psm} (pattern set mining, 16 instances) is given a set of training
items each a vector of Booleans, find some vector which characterizes all
items, or if $k > 1$ find the best $k$ vectors~\cite{guns}; 
\model{photo} (30 instances) is given a set of people and soft constraints
on who they stand next to, place them in a line for a photo; 
\model{rlfap} (radio link frequency assignment, 16 instances) is assigning frequencies to channels with soft constraints that minimize interference~\cite{cabon}; \model{roster} (20 instances) is finding a cyclic roster for a single worker over a number of weeks with soft work pattern constraints; and \model{sugiyama} (5 instances) is a graph layout problem on layered graphs with soft no-edge-crossings constraints~\cite{sugiyama}.

Table~\ref{tab:bounding} shows for each solver: 
`opt' number of instances optimized, 
`s' mean solving time (taking the timeout for instances which timed out), 
`sol' number of instances for which any solution was found, and 
`obj' mean objective (taking the objective from the worst solver for
instances where no objective is available, noting that this may be
too generous).  
The solver which solves the most instances, 
falling back to comparing times and so on, is highlighted.  
\ignore{
For \model{rlfap} we use $\log$-objective due to large variation in objective coefficients between instances.
}

On these problems we see that with standard bound-estimation, we see that unsatisfiable cores is usually better than the unmodified solver but the improvement is not particularly dramatic.  The best version of the unsatisfiable cores algorithm depends on the problem.  Adding notification nearly always improves the nested algorithm, though \model{sugiyama} is an exception which indicates an unusual constraint structure (perhaps there are too many unsatisfiable cores).

If the extra information from unsatisfiable cores is also used for bound-strengthening then dramatic improvements are possible.  The best solver overall is the nested algorithm with notification and (initially empty) LP-based bound strengthening.  We also see good results from the nested or basic algorithm and disjoint-based bound strengthening (without notification).  The trade-off is that disjoint-based bounding is faster but sacrifices pruning power.  Adding notification always helps LP-based strengthening, because the LP receives more information, but never helps disjoint-based strengthening, because overlapping unsatisfiable cores slow down the algorithm without adding strength.

We then tried some much more difficult industrial problems:  
\model{ctt} (curriculum-based timetabling, 32 instances) is finding a weekly
repeating timetable for a university subject to various kinds of soft
availability constraints and soft no-clashes constraints~\cite{digaspero}; 
\model{stein} (Steiner network, 13 instances) is designing a connected
network given a set nodes and arcs with a fixed (building) cost per
arc~\cite{koch};
\model{fcnf} (fixed-charge network flow, 60 instances) is designing a
connected network with fixed (building) and also variable (operating) costs per
arc~\cite{ortega};
and \model{nsp} (nurse scheduling problem, 32 instances) is designing a roster for a hospital ward based on the shift preferences of each individual nurse~\cite{nsplib}.

All of these problems use global constraints taken from \textit{gcc}, \textit{sequence}, \textit{regular} and \textit{network\_flow}.  On \model{ctt} and \model{nsp} we utilize a hybrid search which considers each course/day sequentially, with activity-based search within courses/days.

Furthermore the objective in \model{ctt} and \model{nsp} is not pseudo-Boolean as some $y_i$ are general integer.  Then the unsatisfiable-cores algorithm expresses that e.g.~all nurses should receive their first preference until we have evidence that this should be relaxed to the second preference for certain nurses and so on.  We make the appropriate change to the multiple-decision algorithm, omitted earlier for clarity.  Then unsatisfiable cores e.g.~$\lit{y_1 \ge 3} \vee \lit{y_2 \ge 2}$ say nurse 1 cannot receive 1st or 2nd preference simultaneously with nurse 2 receiving 1st preference.  For bounding we add a linearization e.g.~$1/2 (y_1 - 1) + (y_2 - 1) \ge 1$, into the LP.  We only consider LP-based bounding in this experiment, not disjoint-based.

\begin{table}[t]%
\caption{Evaluating unsatisfiable-core solvers on industrial problems}\label{tab:industrial}%
{\centering \scriptsize \begin{tabular}{|l@{\ }r|r@{}r@{}r@{}r@{}r|r@{}r@{}r@{}r@{}r|r@{}r@{}r@{}r@{}r|r@{}r@{}r@{}r@{}r|}
\hline
\multicolumn{2}{|l|}{\T} & \multicolumn{5}{l|}{fathoming=std} & \multicolumn{5}{l|}{empty LP} & \multicolumn{5}{l|}{redundant LP} & \multicolumn{5}{l|}{LP only} \\
\multicolumn{2}{|l|}{} & \multicolumn{5}{r|}{opt,s,sol,obj,inf} & \multicolumn{5}{r|}{opt,s,sol,obj,inf} & \multicolumn{5}{r|}{opt,s,sol,obj,inf} & \multicolumn{5}{r|}{opt,s,sol,obj,inf} \\
\hline
\T \model{ctt} & b\&b & 0, & 3600s, & 32, & 66843, & 0 & 0, & 3600s, & 32, & 69772, & 0 & 22, & 1332s, & 32, & 82.9, & 0 & 13, & 2734s, & 32, & 86.9, & 0 \\
\multicolumn{2}{|r|}{basic} & 0, & 3600s, & 3, & 62468, & 0 & 0, & 3600s, & 1, & 65611, & 0 & 24, & 1166s, & 28, & 9349, & 0 & 8, & 3059s, & 10, & 48750, & 0 \\
\multicolumn{2}{|r|}{nest} & 0, & 3600s, & 1, & 65485, & 0 & 0, & 3600s, & 1, & 65485, & 0 & {\bf 24}, & {\bf 1067s}, & {\bf 29}, & {\bf 6045}, & {\bf 0} & 7, & 3109s, & 8, & 57222, & 0 \\
\multicolumn{2}{|r|}{nest+not} & 0, & 3600s, & 1, & 65233, & 0 & 0, & 3600s, & 1, & 65233, & 0 & 24, & 1107s, & 29, & 6045, & 0 & 9, & 3094s, & 9, & 51673, & 0 \\
\hline
\T \model{stein} & b\&b & 0, & 3600s, & 13, & 265192, & 0 & 0, & 3600s, & 13, & 9102, & 0 & 2, & 3088s, & 13, & 9001, & 0 & 8, & 1764s, & 10, & 8448, & 0 \\
\multicolumn{2}{|r|}{basic} & 0, & 3600s, & 13, & 8432, & 0 & 3, & 2777s, & 13, & 8533, & 0 & 4, & 2540s, & 13, & 8433, & 0 & 11, & 1041s, & 13, & 3520, & 0 \\
\multicolumn{2}{|r|}{nest} & 0, & 3600s, & 13, & 7657, & 0 & 4, & 2498s, & 13, & 8438, & 0 & 4, & 2503s, & 13, & 6503, & 0 & 11, & 944s, & 13, & 3425, & 0 \\
\multicolumn{2}{|r|}{nest+not} & 1, & 3443s, & 13, & 6957, & 0 & 4, & 2495s, & 13, & 6935, & 0 & 4, & 2507s, & 13, & 5849, & 0 & {\bf 12}, & {\bf 910s}, & {\bf 13}, & {\bf 3422}, & {\bf 0} \\
\hline
\T \model{fcnf} & b\&b & 0, & 3600s, & 2, & 188675, & 0 & 0, & 3600s, & 38, & 185854, & 0 & 9, & 3176s, & 59, & 167292, & 0 & 25, & 2290s, & 47, & 157774, & 0 \\
\multicolumn{2}{|r|}{basic} & 1, & 3556s, & 47, & 171065, & 0 & 0, & 3600s, & 48, & 169844, & 0 & 15, & 2809s, & 60, & 134390, & 0 & 26, & 2180s, & 48, & 153666, & 0 \\
\multicolumn{2}{|r|}{nest} & 1, & 3560s, & 40, & 183401, & 0 & 0, & 3600s, & 45, & 178483, & 0 & 16, & 2722s, & 59, & 135616, & 0 & 23, & 2295s, & 50, & 153992, & 0 \\
\multicolumn{2}{|r|}{nest+not} & 0, & 3600s, & 41, & 181917, & 0 & 7, & 3329s, & 46, & 177765, & 0 & 20, & 2482s, & 60, & 134105, & 0 & {\bf 28}, & {\bf 2009s}, & {\bf 50}, & {\bf 154185}, & {\bf 0} \\
\hline
\T \model{nsp} & b\&b & 3, & 3196s, & 29, & 317, & 1 & 3, & 3177s, & 29, & 319, & 1 & 2, & 3162s, & 30, & 319, & 2 & 1, & 3333s, & 25, & 344, & 2 \\
\multicolumn{2}{|r|}{basic} & 3, & 3263s, & 26, & 278, & 0 & 8, & 2713s, & 26, & 278, & 0 & 8, & 2596s, & 30, & 272, & 2 & 8, & 2537s, & 28, & 272, & 2 \\
\multicolumn{2}{|r|}{nest} & 3, & 3263s, & 27, & 273, & 0 & 8, & 2702s, & 27, & 273, & 0 & 7, & 2591s, & 30, & 267, & 2 & 8, & 2537s, & 28, & 269, & 2 \\
\multicolumn{2}{|r|}{nest+not} & 6, & 3009s, & 27, & 274, & 0 & 8, & 2702s, & 27, & 273, & 0 & {\bf 8}, & {\bf 2512s}, & {\bf 29}, & {\bf 269}, & {\bf 2} & 8, & 2519s, & 27, & 273, & 2 \\
\hline
\end{tabular}
}%
\end{table}%
These problems are extremely difficult to prove infeasible and so
unsatisfiable-cores approaches were not highly successful 
initially (`std'/`empty LP' columns).  We addressed this with a hybrid
CP/MIP approach, by adding a \linear{} decomposition of all CP constraints,
including globals, as a redundant \textit{linear\_program} propagator
(`redundant LP' column).  We also evaluated leaving out the original CP
constraints and using only the LP, so that \chuffed{} becomes similar to a
learning MIP solver such as \scip{} 
using unsatisfiable cores (`LP only' column), but missing advanced
MIP cutting planes and rounding heuristics.

\ignore{
Given that mixed integer programming is a powerful approach to all forms of
combinatorial optimization, and particularly given we are using LP
technology
as part of our CP solver, 
it is valuable to compare with MIP performance on COPPBOs.  
Table~\ref{tab:mip} compares CPLEX and SCIP 
(a learning MIP solver) \pjs{version nos}
versus the best results for our CP solver. We can see on the combinatorial
problems the CP solver is preferable, but the MIP solver dominates on the
industrial problems.  It certainly illustrates that a combination of
unsatisfiable core techniques and MIP/CP hybrid solving is worth pursuing.
}
Table~\ref{tab:industrial} shows the results of these experiments.  We add the column `inf' as number of instances proved infeasible due to hard constraints.  With the enhanced globality due to the \emph{linear\_program} propagator, the unsatisfiable-cores algorithm, in particular the nested algorithm with notification, is a big improvement for CP.  On \model{nsp} (which has preferences), CP is useless without unsatisfiable cores.

\begin{table}[t]%
\caption{Comparing MIP solvers with enhanced CP solver}\label{tab:mip}%
{\centering \scriptsize \begin{tabular}{|@{\ }r|r@{}r@{}r@{}r@{}r|r@{}r@{}r@{}r@{}r|r@{}r@{}r@{}r@{}r|}
\hline
\multicolumn{1}{|l|}{\T} & \multicolumn{5}{l|}{\cplex{}} & \multicolumn{5}{l|}{\scip{} (learning)} & \multicolumn{5}{l|}{\chuffed{} (best solver)} \\
\multicolumn{1}{|l|}{} & \multicolumn{5}{r|}{opt,s,sol,obj,inf} & \multicolumn{5}{r|}{opt,s,sol,obj,inf} & \multicolumn{5}{r|}{opt,s,sol,obj,inf} \\
\hline
\T \model{psm} (k=1) & 9, & 2061s, & 16, & -518, & 0 & 7, & 2514s, & 15, & -447, & 0 & {\bf 12}, & {\bf 966s}, & {\bf 16}, & {\bf -643}, & {\bf 0} \\
\model{psm} (k=2) & 1, & 3375s, & 13, & -565, & 0 & 3, & 3240s, & 8, & -428, & 0 & {\bf 4}, & {\bf 2973s}, & {\bf 16}, & {\bf -602}, & {\bf 0} \\
\model{photo} & 3, & 3488s, & 30, & -14.8, & 0 & 7, & 3216s, & 30, & -15.4, & 0 & {\bf 26}, & {\bf 926s}, & {\bf 30}, & {\bf -16.0}, & {\bf 0} \\
\model{rlfap} & 0, & 3600s, & 5, & 10.4, & 0 & 1, & 3417s, & 8, & 10.2, & 0 & {\bf 3}, & {\bf 3086s}, & {\bf 16}, & {\bf 9.5}, & {\bf 0} \\
\model{roster} & {\bf 18}, & {\bf 365s}, & {\bf 20}, & {\bf 0.5}, & {\bf 0} & 17, & 654s, & 20, & 1.6, & 0 & 16, & 723s, & 16, & 1.6, & 0 \\
\model{sugiyama} & 3, & 1887s, & 5, & 8.8, & 0 & 5, & 816s, & 5, & 8.6, & 0 & {\bf 5}, & {\bf 3s}, & {\bf 5}, & {\bf 8.6}, & {\bf 0} \\
\model{ctt} & {\bf 30}, & {\bf 482s}, & {\bf 32}, & {\bf 81.9}, & {\bf 0} & 21, & 1580s, & 32, & 81.9, & 0 & 24, & 1067s, & 29, & 1474, & 0 \\
\model{stein} & 7, & 1768s, & 13, & 3670, & 0 & 3, & 2770s, & 13, & 3752, & 0 & {\bf 12}, & {\bf 910s}, & {\bf 13}, & {\bf 3422}, & {\bf 0} \\
\model{fcnf} & {\bf 45}, & {\bf 1042s}, & {\bf 60}, & {\bf 111387}, & {\bf 0} & 33, & 1702s, & 60, & 111864, & 0 & 28, & 2009s, & 50, & 142562, & 0 \\
\model{nsp} & {\bf 30}, & {\bf 63s}, & {\bf 30}, & {\bf 263}, & {\bf 2} & 29, & 190s, & 30, & 264, & 2 & 8, & 2512s, & 29, & 267, & 2 \\
\hline
\end{tabular}
}%
\end{table}%
Finally we consider MIP (without unsatisfiable cores) as an alternative to CP for solving soft-constraint problems.  Table~\ref{tab:mip} compare CPLEX 12.0 and SCIP 2.1.1 with the best solver from each of the experiments described above.  On the combinatorial benchmarks CP is clearly superior and much improved with our techniques.  MIP was really only effective on the industrial problems.  Given that timetabling and design problems require globality and that MIP employs many specialized and/or proprietary techniques, MIP should be the first resort for such problems.  On the other hand CP with all our techniques is significantly improved and starts to look competitive, indeed on \model{stein} we beat MIP.  On \model{stein} the `cuts' added for LP-based bounding enforce the connectivity of the network and are similar to~\cite{koch,ortega}, but our approach is much more generic.

\section{Conclusion}%

We have translated unsatisfiable-core methods from MAXSAT to solve 
CP optimization problems with pseudo-Boolean objectives, 
by making use of the facility of LCG solvers to generate unsatisfiable
cores.
This provides one of the first approaches to soft intensionally defined
constraint problems beyond branch and bound that we are aware of, apart from
PBO/WBO~\cite{een,manquinho} which support intensionally-defined \linear{}
constraints only.

To gain the maximum advantage from unsatisfiable cores we needed to extend
the method to generate and use unsatisfiable cores in the middle of search.
This approach to optimization can be substantially better than the
traditional branch and bound approach on intensionally defined optimization
problems.   With our extensions we saw a clear synergy between
\begin{inparaenum}[(i)]\item aggressively assuming that soft constraints hold,
and \item using the resulting information about unsatisfiable or conflicting
soft constraints, for enhanced fathoming\end{inparaenum}.

\bibliography{unsat_core}
\bibliographystyle{splncs03}
\end{document}